\documentclass[pre,aps,twocolumn,floatfix,superscriptaddress]{revtex4-1}
\usepackage{eso-pic,calc}
\usepackage{graphicx}
\usepackage{amsmath, amsthm, amssymb}
\usepackage{epsfig}
\usepackage{bm}
\usepackage{color}

\usepackage[colorlinks=True]{hyperref}  
\hypersetup{
 	colorlinks=true,  
 	linkcolor=blue,  
 	citecolor=blue, 
 	filecolor=magenta,   
 	urlcolor=blue         
 }

\begin{document}
\title{$1/f$ noise in extremal dynamics}
\author{Rahul Chhimpa}
\affiliation{Department of Physics, Institute of Science,  Banaras Hindu University, Varanasi 221 005, India}

\author{Abha Singh}
\affiliation{Department of Physics, Institute of Science,  Banaras Hindu University, Varanasi 221 005, India}

\author{Avinash Chand Yadav\footnote{jnu.avinash@gmail.com}}
\affiliation{Department of Physics, Institute of Science,  Banaras Hindu University, Varanasi 221 005, India}

\begin{abstract}
{The Bak-Sneppen (BS) evolution model remains a well-studied example of self-organized criticality (SOC). We propose a simple variant of the BS model, where the global fitness fluctuations show $1/f^{\alpha}$ noise with a spectral exponent nearly equal to 1 (pink noise). To further corroborate, we compute the two-time autocorrelation function that decays logarithmically. The $1/f$ noise in the global fitness is robust and hyper-universal. We identify the dominance of non-trivial local fitness cross-power spectra.}
\end{abstract}
\maketitle

\section{Introduction}
Scaling features such as fractals in time or $1/f$ noise are ubiquitous~\cite{RevModPhys.53.497}. 
The \emph{$1/f$ noise} has the frequency-dependent scaling in its power spectrum as $1/f^{\alpha}$, with $0<\alpha<2$. While the extreme ends are white noise ($\alpha = 0$) and Brownian noise ($\alpha = 2$), the spectral exponent in many natural systems is typically close to unity (pink noise). The music appears pleasing when it has a balance of predictability and surprise, attributed to the $1/f$ noise present in the intensity changes in music and speech~\cite{VOSS_1975}, including musical rhythms~\cite{doi:10.1073/pnas.1113828109}. Sensory neurons can encode and transmit $1/f$ signals efficiently~\cite{PhysRevLett.94.108103}. The $1/f$ noise also emerges in fossil time series (palaeontological data)~\cite{Ricard_1997}, biological evolution~\cite{PhysRevE.68.031913}, graphene devices~\cite{Balandin_2013}, and the velocity fluctuations in chaotic Hamiltonian dynamics~\cite{PhysRevLett.59.2503}. Surprisingly, the voltage fluctuations in thin film resistors show pink noise over more than six frequency decades~\cite{PhysRevB.27.1233}. The number of accepted spin-flips also exhibits the pink noise in the critical Ising model~\cite{4bpx-t6cq}.
Notice that the pink noise subtly displays a logarithmically decaying correlation function. The logarithmic relaxation~\cite{doi:10.1073/pnas.1120147109} occurs in the electron glass indium oxide~\cite{PhysRevLett.92.066801}, volume relaxation of crumpling paper~\cite{PhysRevLett.88.076101, PhysRevLett.130.258201}, and frictional strength~\cite{Oded_2010}.
Although the ubiquity of $1/f$ noise might suggest a general mechanism, a few common explanations exist. One familiar mechanism is the superposition of exponentially relaxing events with power-law-distributed relaxation times. In particular, if the relaxation time distribution is uniform in an interval, the power spectrum shows pink noise in an intermediate frequency regime.

Despite significant progress, the $1/f$ noise remains an intriguing topic in non-equilibrium statistical physics. 
Bak-Tang-Wiesenfeld (BTW) introduced SOC~\cite{PhysRevLett.59.381, Christensen_2005, MARKOVIC201441, Watkins2016}, a mechanism to explain the $1/f$ noise with simple cellular automata, famously known as the BTW sandpile model. In the BTW model, the avalanche activity recorded at a fast time scale shows $1/f^{\alpha}$ noise with $\alpha \approx 1.6$, and the spectral exponent can be related to avalanche statistics~\cite{Laurson_2005}. Recent work suggests that $\alpha$ can take a value of 1~\cite{shapoval20241varphi} for the mean stress in moderate frequency regimes. Later, several variants of the BTW model were proposed and examined. Maslov-Tang-Zhang (MTZ) studied the BTW model on a narrow strip and found the $1/f$ noise with $\alpha = 1$ in the total mass fluctuations at the drive time scale, and the cutoff frequency grows exponentially with the system size~\cite{PhysRevLett.83.2449}. Interestingly, the MTZ model remains exactly solvable~\cite{PhysRevE.85.061114, Yadav_2022}. A continuous version of the BTW dynamics (the Zhang model) can show the $1/f$ noise with $\alpha = 1$ when driven at the same site~\cite{PhysRevLett.82.472}. In the Zhang model, the cutoff frequency as a function of the system size varies in a power-law or exponential manner for locally conservative and non-conservative dynamics, respectively~\cite{Kumar_2022}. In the locally non-conservative Zhang model, the spectral exponent remains the same in all dimensions (hyper-universal). 

A class of extremal models, such as the BS evolution model~\cite{PhysRevLett.71.4083} and the Robin Hood model~\cite{ZAITSEV1992411}, also exhibit SOC. It is pertinent to recall the BS evolution model. $N$ species are arranged on a circle and characterized by a character \emph{fitness}. Initially, each fitness is assigned randomly from a uniform distribution. In the BS model, the dynamics include reassigning a new random number to the least-fit species and its two nearest neighbors. The BS model respects the Darwinian evolution principle, wherein the least-fit species extincts or mutates, and the interaction includes the coevolution of connected species. In the BS model, several signals have been examined and found to exhibit the $1/f$ noise. Examples include the local activity relevant to return time statistics ($\alpha\approx 0.58$)~\cite{PhysRevLett.73.2162, PhysRevE.53.414}, the number of species below a threshold ($\alpha\approx 1.3$)~\cite{PhysRevE.53.4723,PhysRevE.63.063101, PhysRevE.108.044109}, and the global fitness ($\alpha\approx 1.2$)~\cite{PhysRevE.108.044109}. For the fitness noise, a non-trivial spectral exponent occurs in different variants of the BS model. Such examples are ($\alpha\approx 1.3$) in an anisotropic BS model and ($\alpha\approx 2$) in the random neighbor BS model (mean-field limit)~\cite{PhysRevE.108.044109}. In the BS model, the spectral exponent for the global fitness noise also increases with dimensionality up to the upper critical dimension ${\rm D}_u = 4$~\cite{PhysRevLett.84.2267}.

To the best of our knowledge, in the BS model or its variants, no process has been, so far, reported to have the $1/f$ noise with a spectral exponent $\alpha = 1$, the pink noise, or the canonical case. We suggest evolving only two species (having minimum and maximum fitness) at each time step. Our proposal is a simple modification of the BS model such that the global fitness shows $1/f^{\alpha}$ noise with a spectral exponent value nearly equal to one. To further validate this feature, we compute the two-time autocorrelation function and find logarithmic decay. Also, the variance grows logarithmically with the system size. The $1/f$ noise in global fitness is robust and hyper-universal. We also examine the local fitness power spectra to gain insight into the non-trivial value of the total fitness spectral exponent. The local fitness power spectrum behaves as $\sim 1/\sqrt{N^2f}$ for $N^{-\lambda} \ll f \ll 1/2$. We also find the presence of non-trivial local fitness cross-power spectra, which we attribute to the scaling features in the spatial correlations.

The structure of the paper is as follows: Section~\ref{sec_2} begins with introducing the model. We also show results for the fitness probability density and the probability distribution of distance between consecutive least-fit sites in the critical state. In Sec.~\ref{sec_3}, we present the main results, revealing the long-range temporal correlation of fitness fluctuations by computing power spectra and two-time autocorrelation for both local and total fitness noises. Section~\ref{sec_4} provides a summary.

 \section{An extremal model}{\label{sec_2}}
We consider an ecosystem consisting of $N$ species arranged on the sites of a circle. Each species has an intrinsic variable $\xi$, \emph{fitness}. Initially, we chose fitness as an independent random variable with a uniform distribution between 0 and 1. The species evolve dynamically by selecting the worst and most fit species and replacing their fitness values probabilistically with the same density function. 
 

To motivate the dynamical rules, we assume that naturally extreme fitness values (i.e., maximum and minimum) are unfavorable for survival. 
In biological evolution, the theory of natural selection suggests various forms for the fitness or trait distribution~\cite{cassidy2020, sanjak_2018}. The BS model appears to follow the directional selection, where the mean fitness shifts towards the upper extreme. Our model corroborates the concept of \emph{stabilizing selection}, which disfavors both extreme fitness values, and the distribution tends to be higher and narrower around the mean value.

Other interpretations of the model can include economic systems or games, where species are analogous to agents or players, and fitness is comparable to profit or reward. 
The least profitable economic agent prefers an action likely to yield an enhanced profit. However, if the maximum profit agent overconfidently follows strategies, it is less likely to make an additional gain.
One may argue that if the loser agent has a profit, for example, 0.25, it has a 75\% chance of enhancement after the update. Notice that the updated profit is a uniformly distributed random number, and the loser or winner has a profit less than or greater than 0.5 in the steady state. Similarly, if the winning agent has a profit, for example, 0.75, it has only a 25\% chance of gaining.

\begin{figure}[t]
  \centering
       \scalebox{1.0}{\includegraphics{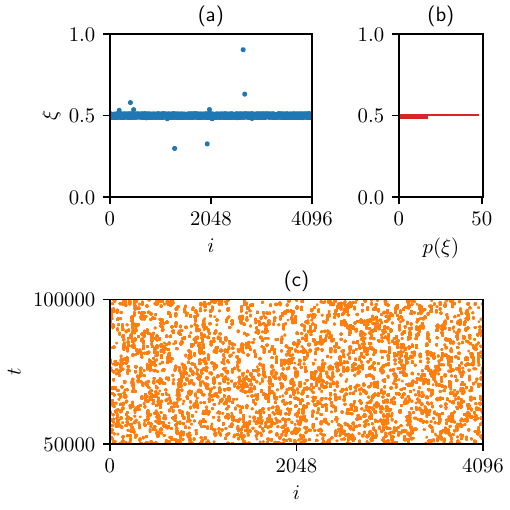}}
          \caption{(a) The fitness profile with $t = 10^5$ and $N = 2^{12}$ and (b) the fitness probability density function. (c) The space-time evolution of the least-fit site.}
  \label{fig_1}
\end{figure}

\begin{figure}[t]
  \centering
       \scalebox{1.0}{\includegraphics{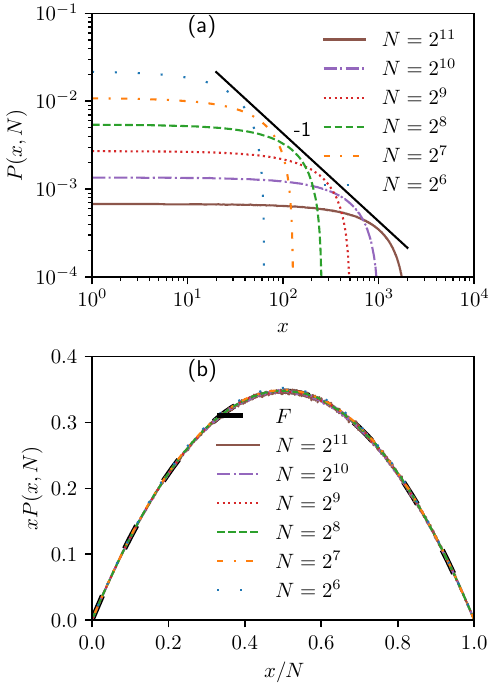}}
          \caption{(a) The probability density $P(x, N)$ for the absolute distance $x$ between two consecutive least-fit sites (in time) with different system sizes $N$. The total events are $10^9$. (b) The probability density follows finite-size scaling behavior [cf. Eq.~(\ref{eq_dist_pdf})]. The thick dashed line represents the theoretically expected scaling function, where the numerical value of the constant is approximately $c \approx 1.39$.}
  \label{fig_ts_ps_lam_1}
\end{figure}

\begin{figure}[t]
  \centering
       \scalebox{1.0}{\includegraphics{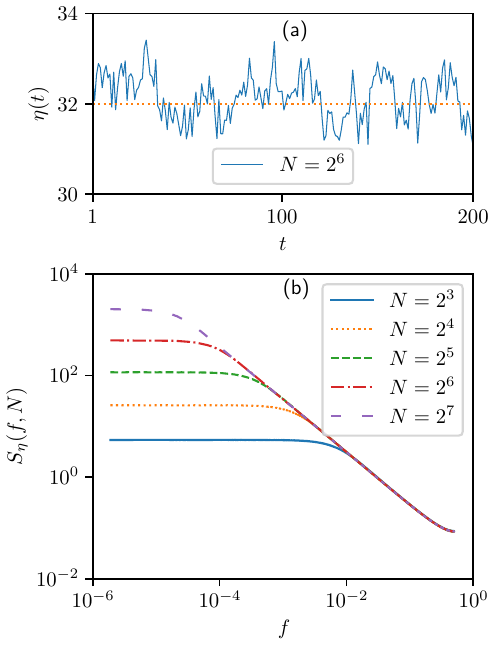}}
          \caption{(a) A typical realization of the global fitness $\eta(t)$. (b) The global fitness power spectra $S_{\eta}(f, N)$ with different system sizes. Signal length is $2^{20}$ and ensemble averaged over $10^5$ independent realizations.}
  \label{fig_gf_ps_1}
\end{figure}

\begin{figure}[t]
  \centering
         \scalebox{1.0}{\includegraphics{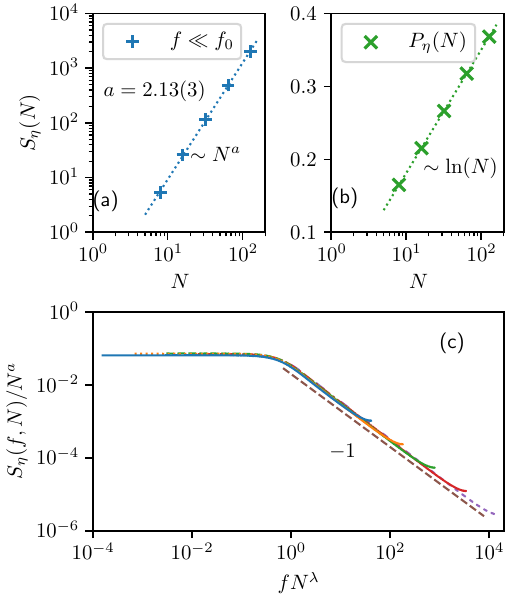}}
       \caption{(a) The system size variation of the global fitness power value in the low frequency component $S_{\eta}(f\ll N^{-\lambda}, N)$. (b) The total power (or variance) of the fitness noise grows logarithmically with increasing system size. (c) The data collapse for Fig.~\ref{fig_gf_ps_1}(b), with $a=2.13$ and $\lambda = 2.13$. The dashed straight line has a slope relating to $\alpha = a/\lambda = 1$.}
  \label{fig_gf_ps_2}
\end{figure}

\begin{figure}[t]
  \centering
       \scalebox{1.0}{\includegraphics{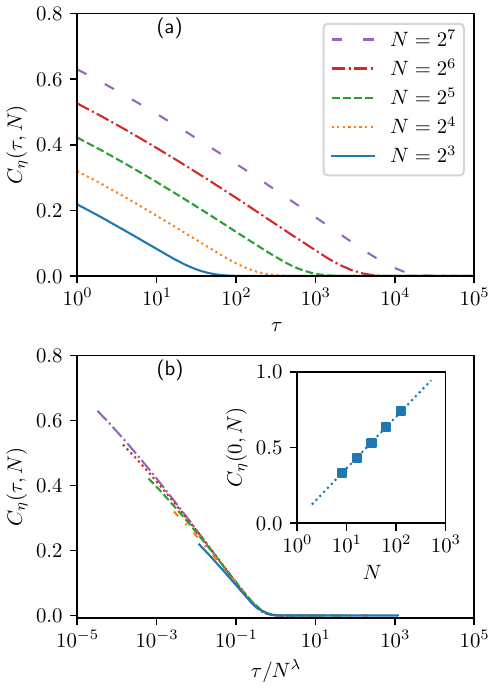}}
          \caption{(a) The two-time autocorrelation function $C_{\eta}(\tau, N)$ for the global fitness $\eta(t)$ with different system sizes $N$. A straight line below the lag time $\tau\ll N^{\lambda}$ is seen on a linear-logarithmic scale. It implies a logarithmically decaying form [cf. Eq.~(\ref{eq_corr})]. We use the lag time $\tau$ running from 0 to $10^5$ and ensemble average over $100$ independent realizations of the signals of length $2^{22}$. (b) The corresponding data collapse as obtained by $\tau \to \tau/N^{\lambda}$ with $\lambda \approx 2.13$. We have shown the correlation function without scaling it by the zero-lag correlation. As shown in the inset, $ C_{\eta}(0, N)$ (or the variance) explicitly depends on the system size and varies in a logarithmic manner. Here the best-fit line is basically: $ C_{\eta}(0, N) \approx 0.34\ln N$. }%$ C_{\eta}(0, N) = 0.340(1)\ln N + 0.021(2)$. }
  \label{fig_corr}
\end{figure}

Notice that the dynamical rules are such that the underlying network topology is irrelevant. The model does not have local interaction. In turn, the model can yield robust and hyper-universal features.
Our extremal model with simple update rules eventually leads to a critical state characterized by intriguing features. As shown in Fig.~\ref{fig_1}, the average fitness value, in the steady state, converges towards $\langle \xi \rangle \to 1/2$ and the fitness probability density becomes a delta function $p(\xi) \sim \delta(\xi -\langle\xi\rangle)$. On the other hand, the fitness density in the BS model behaves as a step function above a threshold fitness $\xi_c \approx 2/3$ as $p(\xi) \sim 1/(1-\xi_c)$ for $\xi_c <\xi <1$~\cite{PhysRevLett.71.4083}, and the mean fitness is $\langle \xi \rangle  = (1+\xi_c)/2$. The space-time evolution of the least fit species remains invariant under space inversion symmetry, and the same site can become active (the least fit) in time recurrently. 

The evolution of the least-fit site resembles a random walk. To uncover the spatial correlation, we examine the probability density of the jump size $x$, the absolute distance between two consecutive least-fit locations in time. We find that the jump size probability decreases linearly but depends on the system size as 
\begin{equation}
P(x, N) \sim \frac{1}{N}\left(1-\frac{x}{N}\right).
\label{eq_dist_pdf}
\end{equation}  
The finite-size scaling (FSS) reveals that we can write Eq.~(\ref{eq_dist_pdf}) as $P(x, N) \sim x^{-1}F(x/N)$, where the scaling function varies as an inverted parabola $F(u) = cu(1-u)$ with $0\le u \le 1$, where $c$ is a constant. Figure~\ref{fig_ts_ps_lam_1} presents numerical results for the jump size probability distribution and its data collapse. On the other hand, in the BS model, the jump size probability decays in a power-law manner with a critical exponent of approximately 3~\cite{PhysRevLett.71.4083}. 

The dynamics mimic only the effects of external perturbations, and there is no internal interaction (non-interacting).
The change in the least-fit species represents an effort to achieve enhancement (more likely to push upward), while the update in the most-fit species is like a suppression effect (more likely to pull down). The two competing forces (enhancing or suppressing fitness or survivability) lead the system to self-organize, and eventually, the system displays intriguing scaling features leading to pink noise (shown below). 

\begin{figure}[t]
  \centering
       \scalebox{1.0}{\includegraphics{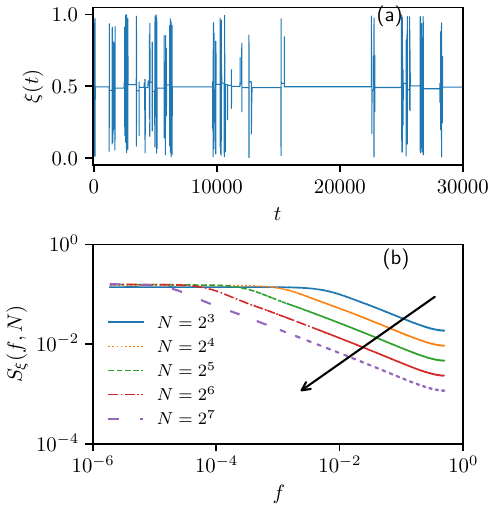}}
          \caption{(a) A portion of the local fitness time series $\xi(t)$. (b) The local fitness power spectra for different system sizes. The arrow indicates the effect of increasing system size values.}
  \label{fig_ts_ps_lf_1}
\end{figure}

\begin{figure}[t]
  \centering
         \scalebox{1.0}{\includegraphics{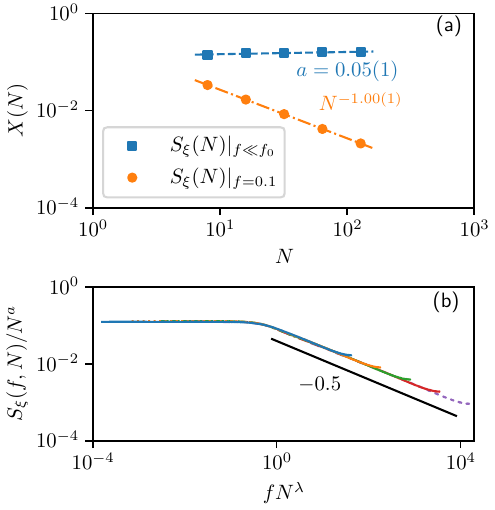}}
       \caption{(a) The system size scaling of the power in the low-frequency component and at a frequency above the cutoff frequency. (b) The data collapse of the power spectra [cf. Fig.~\ref{fig_ts_ps_lf_1}(b)].}%, with $a = 0.05$ and $\lambda = 2.13$. }%: $u \sim fN^{\lambda}$ and $H(u) \sim S(f, N)/N^{a}$.}
  \label{fig_ts_ps_lf_2}
\end{figure}

\section{Simulation results for the temporal correlation of fitness noise}{\label{sec_3}} 
We, here, investigate the temporal correlations in global fitness, $\eta(t) = \sum_{i=1}^{N} \xi_i(t)$, by computing first the power spectra for different system sizes and applying the FSS method to extract the critical exponents and scaling function. A typical signal generated via Monte Carlo simulation is in Fig.~\ref{fig_gf_ps_1}(a). Figure~\ref{fig_gf_ps_1}(b) shows the global fitness power spectra for different system sizes $N$. We employed the standard fast Fourier transform algorithm to compute the power spectrum: $S_{\eta}(f) =  \lim_{T\to \infty} \langle |\tilde{\eta} (f)|^2\rangle/T$, where $T$ is the signal length and $\langle \cdot \rangle$ denotes ensemble average. Figures~\ref{fig_gf_ps_2}(a) and (b) show the system size dependence of the power in the low-frequency component $S_{\eta}(N, f\ll f_0) \sim N^{a}$ and the total power $P_{\eta}(N)$, respectively. The FSS analysis, as shown in Fig.~\ref{fig_gf_ps_2}(c), clearly suggests that the global fitness exhibits $1/f^{\alpha}$ noise with $\alpha = 1$ in a frequency regime of $f_0 \ll f \ll 1/2$ [cf. Fig.~\ref{fig_gf_ps_1}(b)]. Notice that $S_{\eta}(f, N) \sim N^aH_{\eta}(fN^{\lambda})$, where the scaling function varies as $H_{\eta}(u) \sim u^{-\alpha}$ for $u \gg 1$ and constant for $u \ll 1$. In the frequency regime $f\gg N^{-\lambda}$, the system size independence of the power spectrum yields a scaling relation $\alpha = a/\lambda$. The cutoff frequency scales with the system size as $f_0 \sim N^{-\lambda}$ with $\lambda \sim 2.13$. Interestingly, the total power grows logarithmically with the system size [cf. Fig.~\ref{fig_gf_ps_2}(b)].

We also examine the two-time autocorrelation function for different system sizes to validate our striking observation of the $1/f^{\alpha}$ noise with $\alpha = 1$ for the global fitness. As shown in Fig.~\ref{fig_corr}(a), we find that the correlation function decays in a logarithmic manner. Applying the FSS method, we can similarly obtain data collapse [cf. Fig.~\ref{fig_corr}(b)] that suggests
\begin{equation}
C_{\eta}(\tau, N) - C_{\eta}(0, N) \sim \begin{cases}-\ln|\tau/N^{\lambda}|,~~{\rm for}~~ \tau\ll N^{\lambda}, \\ 0,~~~~~~~~~~~~~~~~{\rm for}~~ \tau\gg N^{\lambda},\end{cases}
\label{eq_corr}
\end{equation}   
where the zero lag correlation or the variance grows logarithmically: $C_{\eta}(0, N) \sim \ln N$ [cf. inset in Fig.~\ref{fig_corr}(b)]. Since the power spectrum of a wide-sense stationarity process is simply the Fourier transformation of the two-time autocorrelation function (Wiener-Khinchin theorem), such a logarithmically decaying correlation function [cf. Eq.~(\ref{eq_corr})] implies that the underlying noise has a power spectrum of the asymptotic form $1/f^{\alpha}$ with $\alpha = 1$. 

To gain a better understanding of the non-trivial spectral exponent value of the global fitness power spectra, we also study the local fitness power spectra with different system sizes [cf. Figs.~\ref{fig_ts_ps_lf_1}-\ref{fig_ts_ps_lf_2}] and find that 
\begin{equation}
S_{\xi}(f, N) \sim \begin{cases}{\rm constant}, ~~~f\ll f_0 \\ 1/\sqrt{fN^2}, ~~~f_0 \ll f \ll 1/2. \end{cases}
\label{eq_ps_local}
\end{equation}
The scaling feature of the local fitness power spectrum suggests that the underlying local fitness correlation function decays slowly in an algebraic form $\sim 1/\sqrt{\tau}$. We also note that the power spectrum shown in Eq.~(\ref{eq_ps_local}) remains the same for each site. The explicit system size dependence in the non-trivial frequency regime eventually modifies the spectral exponent $\alpha = (a+1)/\lambda$. The spectral exponent $\alpha \approx 1/2$ for both the local fitness and the local activity (not shown) suggests that the underlying dynamical wandering of the least fit site is equivalent to a simple random walk. Interestingly, the spin noise in semiconductor nanowires~\cite{PhysRevLett.107.156602} can show a power spectrum of the form $1/\sqrt{f}$ (slow spin relaxation) for different regimes of electron transport and spin dynamics. 

\begin{figure}[t]
	\centering
	\scalebox{1.0}{\includegraphics{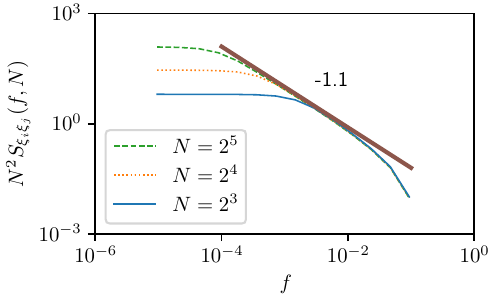}}
	\caption{The log-binned local fitness cross-power spectra with different system sizes. The straight line has a slope of 1.1.}
	\label{fig_ts_ps_clf_3}
\end{figure}

A trivial sum of the local fitness power spectra over all sites does not yield the global fitness power spectrum with $\alpha = 1$. It implies the existence of a non-trivial local cross-power spectrum emerging because of scaling features present in the spatial correlation, as noted from the random walk jump-size probability density. We have also examined the local fitness cross-power spectrum $S_{\xi_i\xi_j}(f, N, r)$ [cf. Fig.~\ref{fig_ts_ps_clf_3}], where $r$ is the distance between the two sites where we record the local fitness signals $\xi_i(t)$ and $\xi_j(t)$. We find the local fitness cross-power spectrum does not depend on the distance $r$ but shows a $1/f^{\alpha}$ type scaling with $\alpha \sim 1.1$ along with a system size dependence as
\begin{equation}
S_{\xi_i\xi_j}(f, N) \sim \begin{cases}{\rm constant}, ~~~~~~f\ll f_0 \\ 1/(f^{\alpha}N^2), ~~~f_0 \ll f \ll 1/2. \end{cases}
\label{eq_ps_local_cross}
\end{equation} 

It seems that the local fitness cross-power spectrum plays a dominant role in the global fitness power spectrum. One can easily note that the total fitness two-time autocorrelation function in terms of the local correlations is 
\begin{equation}
C_{\eta}(\tau, N) = \langle \eta(t)\eta(t+\tau)\rangle = \sum_{i, j}^{N}  \langle \xi_i(t)\xi_j(t+\tau)\rangle.\nonumber
\end{equation}
Separating the two-time autocorrelation and cross-correlations (local), we can write
\begin{equation}
C_{\eta}(\tau, N) =  \sum_{i=j}^{N}  \langle \xi_i(t)\xi_j(t+\tau)\rangle + \sum_{i\neq j}^{N}  \langle \xi_i(t)\xi_j(t+\tau)\rangle. \nonumber
\end{equation} 
For $\tau\ll N^{\lambda}$, the global fitness two-time autocorrelation behaves as [cf. Eqs.~(\ref{eq_ps_local}) and (\ref{eq_ps_local_cross})]
\begin{equation}
C_{\eta}(0, N) - C_{\eta}(\tau, L) \approx A_1/\sqrt{|\tau/N^{\lambda}|} +   A_2\ln{|\tau/N^{\lambda}|}, 
\label{eq_corr_1}
\end{equation} 
with $A_2/A_1\gg 1$. The first term in Eq.~(\ref{eq_corr_1}) represents the all return time probability of a simple random walk~\cite{PhysRevE.53.414}.

\section{Summary}{\label{sec_4}}
To summarize, we have provided a simple self-organized extremal model (a variant of the Bak-Sneppen evolution dynamics) that demonstrates that the evolution of species only with extremal (minimum and maximum) fitness values in an evolutionary system can lead to striking scaling behavior. The macroscopic noise (global fitness) shows a signature of pink noise, the canonical case of the $1/f^{\alpha}$ noise with $\alpha \approx 1$, which we confirmed by computing the two-time autocorrelation function that decays in a logarithmic manner. Similarly, the variance, or total power, grows logarithmically with the system size. A lower cutoff frequency exists and scales with system size $\sim N^{-2.1}$. The emergent $1/f$ noise is indeed robust and hyper-universal. The microscopic noise (local fitness) displays a slow relaxation $\sim 1/\sqrt{\tau}$, or the corresponding power spectrum varies as $1/\sqrt{f}$. However, a non-trivial local cross-power spectrum $\sim 1/f^{1.1}$ emerges as a dominant contribution to the total fitness fluctuations. Although the dynamics are non-interacting, the two competing external perturbations — enhancement and suppression effects — eventually lead the system to self-organize into a critical state, exhibiting intriguing scaling features. 

\section*{ACKNOWLEDGMENTS}
RC and AS acknowledge financial support from the Senior Research Fellowship, UGC, India, and Banaras Hindu University (Grant No. R/Dev./Sch/UGC Non-Net Fello./2022-23/53315), respectively. 

\bibliography{s1sources}

\begin{thebibliography}{37}%
\makeatletter
\providecommand \@ifxundefined [1]{%
 \@ifx{#1\undefined}
}%
\providecommand \@ifnum [1]{%
 \ifnum #1\expandafter \@firstoftwo
 \else \expandafter \@secondoftwo
 \fi
}%
\providecommand \@ifx [1]{%
 \ifx #1\expandafter \@firstoftwo
 \else \expandafter \@secondoftwo
 \fi
}%
\providecommand \natexlab [1]{#1}%
\providecommand \enquote  [1]{``#1''}%
\providecommand \bibnamefont  [1]{#1}%
\providecommand \bibfnamefont [1]{#1}%
\providecommand \citenamefont [1]{#1}%
\providecommand \href@noop [0]{\@secondoftwo}%
\providecommand \href [0]{\begingroup \@sanitize@url \@href}%
\providecommand \@href[1]{\@@startlink{#1}\@@href}%
\providecommand \@@href[1]{\endgroup#1\@@endlink}%
\providecommand \@sanitize@url [0]{\catcode `\\12\catcode `\$12\catcode
  `\&12\catcode `\#12\catcode `\^12\catcode `\_12\catcode `\%12\relax}%
\providecommand \@@startlink[1]{}%
\providecommand \@@endlink[0]{}%
\providecommand \url  [0]{\begingroup\@sanitize@url \@url }%
\providecommand \@url [1]{\endgroup\@href {#1}{\urlprefix }}%
\providecommand \urlprefix  [0]{URL }%
\providecommand \Eprint [0]{\href }%
\providecommand \doibase [0]{http://dx.doi.org/}%
\providecommand \selectlanguage [0]{\@gobble}%
\providecommand \bibinfo  [0]{\@secondoftwo}%
\providecommand \bibfield  [0]{\@secondoftwo}%
\providecommand \translation [1]{[#1]}%
\providecommand \BibitemOpen [0]{}%
\providecommand \bibitemStop [0]{}%
\providecommand \bibitemNoStop [0]{.\EOS\space}%
\providecommand \EOS [0]{\spacefactor3000\relax}%
\providecommand \BibitemShut  [1]{\csname bibitem#1\endcsname}%
\let\auto@bib@innerbib\@empty
%</preamble>
\bibitem [{\citenamefont {Dutta}\ and\ \citenamefont
  {Horn}(1981)}]{RevModPhys.53.497}%
  \BibitemOpen
  \bibfield  {author} {\bibinfo {author} {\bibfnamefont {P.}~\bibnamefont
  {Dutta}}\ and\ \bibinfo {author} {\bibfnamefont {P.~M.}\ \bibnamefont
  {Horn}},\ }{\it \bibinfo {title} {Low-frequency fluctuations in solids:
  $\frac{1}{f}$ noise},}\ \href {\doibase 10.1103/RevModPhys.53.497} {\bibfield
   {journal} {\bibinfo  {journal} {Rev. Mod. Phys.}\ }\textbf {\bibinfo
  {volume} {53}},\ \bibinfo {pages} {497} (\bibinfo {year} {1981})}\BibitemShut
  {NoStop}%
\bibitem [{\citenamefont {Voss}\ and\ \citenamefont
  {Clarke}(1975)}]{VOSS_1975}%
  \BibitemOpen
  \bibfield  {author} {\bibinfo {author} {\bibfnamefont {R.~F.}\ \bibnamefont
  {Voss}}\ and\ \bibinfo {author} {\bibfnamefont {J.}~\bibnamefont {Clarke}},\
  }{\it \bibinfo {title} {‘1/f noise’ in music and speech},}\ \href
  {\doibase https://doi.org/10.1038/258317a0} {\bibfield  {journal} {\bibinfo
  {journal} {Nature (London)}\ }\textbf {\bibinfo {volume} {258}},\ \bibinfo
  {pages} {317} (\bibinfo {year} {1975})}\BibitemShut {NoStop}%
\bibitem [{\citenamefont {Levitin}\ \emph {et~al.}(2012)\citenamefont
  {Levitin}, \citenamefont {Chordia},\ and\ \citenamefont
  {Menon}}]{doi:10.1073/pnas.1113828109}%
  \BibitemOpen
  \bibfield  {author} {\bibinfo {author} {\bibfnamefont {D.~J.}\ \bibnamefont
  {Levitin}}, \bibinfo {author} {\bibfnamefont {P.}~\bibnamefont {Chordia}}, \
  and\ \bibinfo {author} {\bibfnamefont {V.}~\bibnamefont {Menon}},\ }{\it
  \bibinfo {title} {Musical rhythm spectra from Bach to Joplin obey a $1/f$
  power law},}\ \href {\doibase 10.1073/pnas.1113828109} {\bibfield  {journal}
  {\bibinfo  {journal} {Proc. Natl. Acad. Sci. USA}\ }\textbf {\bibinfo
  {volume} {109}},\ \bibinfo {pages} {3716} (\bibinfo {year}
  {2012})}\BibitemShut {NoStop}%
\bibitem [{\citenamefont {Yu}\ \emph {et~al.}(2005)\citenamefont {Yu},
  \citenamefont {Romero},\ and\ \citenamefont {Lee}}]{PhysRevLett.94.108103}%
  \BibitemOpen
  \bibfield  {author} {\bibinfo {author} {\bibfnamefont {Y.}~\bibnamefont
  {Yu}}, \bibinfo {author} {\bibfnamefont {R.}~\bibnamefont {Romero}}, \ and\
  \bibinfo {author} {\bibfnamefont {T.~S.}\ \bibnamefont {Lee}},\ }{\it
  \bibinfo {title} {Preference of Sensory Neural Coding for $1/f$ Signals},}\
  \href {\doibase 10.1103/PhysRevLett.94.108103} {\bibfield  {journal}
  {\bibinfo  {journal} {Phys. Rev. Lett.}\ }\textbf {\bibinfo {volume} {94}},\
  \bibinfo {pages} {108103} (\bibinfo {year} {2005})}\BibitemShut {NoStop}%
\bibitem [{\citenamefont {Sol\'e}\ \emph {et~al.}(1997)\citenamefont {Sol\'e},
  \citenamefont {Manrubia}, \citenamefont {Benton},\ and\ \citenamefont
  {Bak}}]{Ricard_1997}%
  \BibitemOpen
  \bibfield  {author} {\bibinfo {author} {\bibfnamefont {R.~V.}\ \bibnamefont
  {Sol\'e}}, \bibinfo {author} {\bibfnamefont {S.~C.}\ \bibnamefont
  {Manrubia}}, \bibinfo {author} {\bibfnamefont {M.}~\bibnamefont {Benton}}, \
  and\ \bibinfo {author} {\bibfnamefont {P.}~\bibnamefont {Bak}},\ }{\it
  \bibinfo {title} {Self-similarity of extinction statistics in the fossil
  record},}\ \href {\doibase https://doi.org/10.1038/41996} {\bibfield
  {journal} {\bibinfo  {journal} {Nature (London)}\ }\textbf {\bibinfo {volume}
  {388}},\ \bibinfo {pages} {764} (\bibinfo {year} {1997})}\BibitemShut
  {NoStop}%
\bibitem [{\citenamefont {Rikvold}\ and\ \citenamefont
  {Zia}(2003)}]{PhysRevE.68.031913}%
  \BibitemOpen
  \bibfield  {author} {\bibinfo {author} {\bibfnamefont {P.~A.}\ \bibnamefont
  {Rikvold}}\ and\ \bibinfo {author} {\bibfnamefont {R.~K.~P.}\ \bibnamefont
  {Zia}},\ }{\it \bibinfo {title} {Punctuated equilibria and $1/f$ noise in a
  biological coevolution model with individual-based dynamics},}\ \href
  {\doibase 10.1103/PhysRevE.68.031913} {\bibfield  {journal} {\bibinfo
  {journal} {Phys. Rev. E}\ }\textbf {\bibinfo {volume} {68}},\ \bibinfo
  {pages} {031913} (\bibinfo {year} {2003})}\BibitemShut {NoStop}%
\bibitem [{\citenamefont {Balandin}(2013)}]{Balandin_2013}%
  \BibitemOpen
  \bibfield  {author} {\bibinfo {author} {\bibfnamefont {A.~A.}\ \bibnamefont
  {Balandin}},\ }{\it \bibinfo {title} {Low-frequency 1/f noise in graphene
  devices},}\ \href {\doibase https://doi.org/10.1038/nnano.2013.144}
  {\bibfield  {journal} {\bibinfo  {journal} {Nat. Nanotechnol.}\ }\textbf
  {\bibinfo {volume} {8}},\ \bibinfo {pages} {549} (\bibinfo {year}
  {2013})}\BibitemShut {NoStop}%
\bibitem [{\citenamefont {Geisel}\ \emph {et~al.}(1987)\citenamefont {Geisel},
  \citenamefont {Zacherl},\ and\ \citenamefont {Radons}}]{PhysRevLett.59.2503}%
  \BibitemOpen
  \bibfield  {author} {\bibinfo {author} {\bibfnamefont {T.}~\bibnamefont
  {Geisel}}, \bibinfo {author} {\bibfnamefont {A.}~\bibnamefont {Zacherl}}, \
  and\ \bibinfo {author} {\bibfnamefont {G.}~\bibnamefont {Radons}},\ }{\it
  \bibinfo {title} {Generic $\frac{1}{f}$ Noise in Chaotic Hamiltonian
  Dynamics},}\ \href {\doibase 10.1103/PhysRevLett.59.2503} {\bibfield
  {journal} {\bibinfo  {journal} {Phys. Rev. Lett.}\ }\textbf {\bibinfo
  {volume} {59}},\ \bibinfo {pages} {2503} (\bibinfo {year}
  {1987})}\BibitemShut {NoStop}%
\bibitem [{\citenamefont {Pellegrini}\ \emph {et~al.}(1983)\citenamefont
  {Pellegrini}, \citenamefont {Saletti}, \citenamefont {Terreni},\ and\
  \citenamefont {Prudenziati}}]{PhysRevB.27.1233}%
  \BibitemOpen
  \bibfield  {author} {\bibinfo {author} {\bibfnamefont {B.}~\bibnamefont
  {Pellegrini}}, \bibinfo {author} {\bibfnamefont {R.}~\bibnamefont {Saletti}},
  \bibinfo {author} {\bibfnamefont {P.}~\bibnamefont {Terreni}}, \ and\
  \bibinfo {author} {\bibfnamefont {M.}~\bibnamefont {Prudenziati}},\ }{\it
  \bibinfo {title} {$\frac{1}{{f}^{\ensuremath{\gamma}}}$ noise in thick-film
  resistors as an effect of tunnel and thermally activated emissions, from
  measures versus frequency and temperature},}\ \href {\doibase
  10.1103/PhysRevB.27.1233} {\bibfield  {journal} {\bibinfo  {journal} {Phys.
  Rev. B}\ }\textbf {\bibinfo {volume} {27}},\ \bibinfo {pages} {1233}
  (\bibinfo {year} {1983})}\BibitemShut {NoStop}%
\bibitem [{\citenamefont {Chhimpa}\ and\ \citenamefont
  {Yadav}(2025)}]{4bpx-t6cq}%
  \BibitemOpen
  \bibfield  {author} {\bibinfo {author} {\bibfnamefont {R.}~\bibnamefont
  {Chhimpa}}\ and\ \bibinfo {author} {\bibfnamefont {A.~C.}\ \bibnamefont
  {Yadav}},\ }{\it \bibinfo {title} {$1/f$ noise in the Ising model},}\ \href
  {\doibase 10.1103/4bpx-t6cq} {\bibfield  {journal} {\bibinfo  {journal}
  {Phys. Rev. E}\ }\textbf {\bibinfo {volume} {111}},\ \bibinfo {pages}
  {064130} (\bibinfo {year} {2025})}\BibitemShut {NoStop}%
\bibitem [{\citenamefont {Amir}\ \emph {et~al.}(2012)\citenamefont {Amir},
  \citenamefont {Oreg},\ and\ \citenamefont
  {Imry}}]{doi:10.1073/pnas.1120147109}%
  \BibitemOpen
  \bibfield  {author} {\bibinfo {author} {\bibfnamefont {A.}~\bibnamefont
  {Amir}}, \bibinfo {author} {\bibfnamefont {Y.}~\bibnamefont {Oreg}}, \ and\
  \bibinfo {author} {\bibfnamefont {Y.}~\bibnamefont {Imry}},\ }{\it \bibinfo
  {title} {On relaxations and aging of various glasses},}\ \href {\doibase
  10.1073/pnas.1120147109} {\bibfield  {journal} {\bibinfo  {journal} {Proc.
  Natl. Acad. Sci. USA}\ }\textbf {\bibinfo {volume} {109}},\ \bibinfo {pages}
  {1850} (\bibinfo {year} {2012})}\BibitemShut {NoStop}%
\bibitem [{\citenamefont {Orlyanchik}\ and\ \citenamefont
  {Ovadyahu}(2004)}]{PhysRevLett.92.066801}%
  \BibitemOpen
  \bibfield  {author} {\bibinfo {author} {\bibfnamefont {V.}~\bibnamefont
  {Orlyanchik}}\ and\ \bibinfo {author} {\bibfnamefont {Z.}~\bibnamefont
  {Ovadyahu}},\ }{\it \bibinfo {title} {Stress Aging in the Electron Glass},}\
  \href {\doibase 10.1103/PhysRevLett.92.066801} {\bibfield  {journal}
  {\bibinfo  {journal} {Phys. Rev. Lett.}\ }\textbf {\bibinfo {volume} {92}},\
  \bibinfo {pages} {066801} (\bibinfo {year} {2004})}\BibitemShut {NoStop}%
\bibitem [{\citenamefont {Matan}\ \emph {et~al.}(2002)\citenamefont {Matan},
  \citenamefont {Williams}, \citenamefont {Witten},\ and\ \citenamefont
  {Nagel}}]{PhysRevLett.88.076101}%
  \BibitemOpen
  \bibfield  {author} {\bibinfo {author} {\bibfnamefont {K.}~\bibnamefont
  {Matan}}, \bibinfo {author} {\bibfnamefont {R.~B.}\ \bibnamefont {Williams}},
  \bibinfo {author} {\bibfnamefont {T.~A.}\ \bibnamefont {Witten}}, \ and\
  \bibinfo {author} {\bibfnamefont {S.~R.}\ \bibnamefont {Nagel}},\ }{\it
  \bibinfo {title} {Crumpling a Thin Sheet},}\ \href {\doibase
  10.1103/PhysRevLett.88.076101} {\bibfield  {journal} {\bibinfo  {journal}
  {Phys. Rev. Lett.}\ }\textbf {\bibinfo {volume} {88}},\ \bibinfo {pages}
  {076101} (\bibinfo {year} {2002})}\BibitemShut {NoStop}%
\bibitem [{\citenamefont {Lahini}\ \emph {et~al.}(2023)\citenamefont {Lahini},
  \citenamefont {Rubinstein},\ and\ \citenamefont
  {Amir}}]{PhysRevLett.130.258201}%
  \BibitemOpen
  \bibfield  {author} {\bibinfo {author} {\bibfnamefont {Y.}~\bibnamefont
  {Lahini}}, \bibinfo {author} {\bibfnamefont {S.~M.}\ \bibnamefont
  {Rubinstein}}, \ and\ \bibinfo {author} {\bibfnamefont {A.}~\bibnamefont
  {Amir}},\ }{\it \bibinfo {title} {Crackling Noise during Slow Relaxations in
  Crumpled Sheets},}\ \href {\doibase 10.1103/PhysRevLett.130.258201}
  {\bibfield  {journal} {\bibinfo  {journal} {Phys. Rev. Lett.}\ }\textbf
  {\bibinfo {volume} {130}},\ \bibinfo {pages} {258201} (\bibinfo {year}
  {2023})}\BibitemShut {NoStop}%
\bibitem [{\citenamefont {Ben-David}\ \emph {et~al.}(2010)\citenamefont
  {Ben-David}, \citenamefont {Rubinstein},\ and\ \citenamefont
  {Fineberg}}]{Oded_2010}%
  \BibitemOpen
  \bibfield  {author} {\bibinfo {author} {\bibfnamefont {O.}~\bibnamefont
  {Ben-David}}, \bibinfo {author} {\bibfnamefont {S.~M.}\ \bibnamefont
  {Rubinstein}}, \ and\ \bibinfo {author} {\bibfnamefont {J.}~\bibnamefont
  {Fineberg}},\ }{\it \bibinfo {title} {Slip-stick and the evolution of
  frictional strength},}\ \href {\doibase https://doi.org/10.1038/nature08676}
  {\bibfield  {journal} {\bibinfo  {journal} {Nature (London)}\ }\textbf
  {\bibinfo {volume} {463}},\ \bibinfo {pages} {76} (\bibinfo {year}
  {2010})}\BibitemShut {NoStop}%
\bibitem [{\citenamefont {Bak}\ \emph {et~al.}(1987)\citenamefont {Bak},
  \citenamefont {Tang},\ and\ \citenamefont {Wiesenfeld}}]{PhysRevLett.59.381}%
  \BibitemOpen
  \bibfield  {author} {\bibinfo {author} {\bibfnamefont {P.}~\bibnamefont
  {Bak}}, \bibinfo {author} {\bibfnamefont {C.}~\bibnamefont {Tang}}, \ and\
  \bibinfo {author} {\bibfnamefont {K.}~\bibnamefont {Wiesenfeld}},\ }{\it
  \bibinfo {title} {Self-organized criticality: An explanation of the 1/f
  noise},}\ \href {\doibase 10.1103/PhysRevLett.59.381} {\bibfield  {journal}
  {\bibinfo  {journal} {Phys. Rev. Lett.}\ }\textbf {\bibinfo {volume} {59}},\
  \bibinfo {pages} {381} (\bibinfo {year} {1987})}\BibitemShut {NoStop}%
\bibitem [{\citenamefont {Christensen}\ and\ \citenamefont
  {Moloney}(2005)}]{Christensen_2005}%
  \BibitemOpen
  \bibfield  {author} {\bibinfo {author} {\bibfnamefont {K.}~\bibnamefont
  {Christensen}}\ and\ \bibinfo {author} {\bibfnamefont {N.~R.}\ \bibnamefont
  {Moloney}},\ }\href@noop {} {\emph {\bibinfo {title} {Complexity and
  Criticality}}}\ (\bibinfo  {publisher} {Imperial College Press, London},\
  \bibinfo {year} {2005})\BibitemShut {NoStop}%
\bibitem [{\citenamefont {Markovi\'c}\ and\ \citenamefont
  {Gros}(2014)}]{MARKOVIC201441}%
  \BibitemOpen
  \bibfield  {author} {\bibinfo {author} {\bibfnamefont {D.}~\bibnamefont
  {Markovi\'c}}\ and\ \bibinfo {author} {\bibfnamefont {C.}~\bibnamefont
  {Gros}},\ }{\it \bibinfo {title} {Power laws and self-organized criticality
  in theory and nature},}\ \href {\doibase
  https://doi.org/10.1016/j.physrep.2013.11.002} {\bibfield  {journal}
  {\bibinfo  {journal} {Phys. Rep.}\ }\textbf {\bibinfo {volume} {536}},\
  \bibinfo {pages} {41} (\bibinfo {year} {2014})}\BibitemShut {NoStop}%
\bibitem [{\citenamefont {Watkins}\ \emph {et~al.}(2016)\citenamefont
  {Watkins}, \citenamefont {Pruessner}, \citenamefont {Chapman}, \citenamefont
  {Crosby},\ and\ \citenamefont {Jensen}}]{Watkins2016}%
  \BibitemOpen
  \bibfield  {author} {\bibinfo {author} {\bibfnamefont {N.~W.}\ \bibnamefont
  {Watkins}}, \bibinfo {author} {\bibfnamefont {G.}~\bibnamefont {Pruessner}},
  \bibinfo {author} {\bibfnamefont {S.~C.}\ \bibnamefont {Chapman}}, \bibinfo
  {author} {\bibfnamefont {N.~B.}\ \bibnamefont {Crosby}}, \ and\ \bibinfo
  {author} {\bibfnamefont {H.~J.}\ \bibnamefont {Jensen}},\ }{\it \bibinfo
  {title} {25 Years of Self-organized Criticality: Concepts and
  Controversies},}\ \href {\doibase 10.1007/s11214-015-0155-x} {\bibfield
  {journal} {\bibinfo  {journal} {Space. Sci. Rev.}\ }\textbf {\bibinfo
  {volume} {198}},\ \bibinfo {pages} {3} (\bibinfo {year} {2016})}\BibitemShut
  {NoStop}%
\bibitem [{\citenamefont {Laurson}\ \emph {et~al.}(2005)\citenamefont
  {Laurson}, \citenamefont {Alava},\ and\ \citenamefont
  {Zapperi}}]{Laurson_2005}%
  \BibitemOpen
  \bibfield  {author} {\bibinfo {author} {\bibfnamefont {L.}~\bibnamefont
  {Laurson}}, \bibinfo {author} {\bibfnamefont {M.~J.}\ \bibnamefont {Alava}},
  \ and\ \bibinfo {author} {\bibfnamefont {S.}~\bibnamefont {Zapperi}},\ }{\it
  \bibinfo {title} {Power spectra of self-organized critical sandpiles},}\
  \href {\doibase 10.1088/1742-5468/2005/11/L11001} {\bibfield  {journal}
  {\bibinfo  {journal} {J. Stat. Mech.}\ }\textbf {\bibinfo {volume} {2005}},\
  \bibinfo {pages} {L11001} (\bibinfo {year} {2005})}\BibitemShut {NoStop}%
\bibitem [{\citenamefont {Shapoval}\ and\ \citenamefont
  {Shnirman}(2024)}]{shapoval20241varphi}%
  \BibitemOpen
  \bibfield  {author} {\bibinfo {author} {\bibfnamefont {A.}~\bibnamefont
  {Shapoval}}\ and\ \bibinfo {author} {\bibfnamefont {M.}~\bibnamefont
  {Shnirman}},\ }{\it \bibinfo {title} {Explanation of flicker noise with the
  Bak-Tang-Wiesenfeld model of self-organized criticality},}\ \href {\doibase
  10.1103/PhysRevE.110.014106} {\bibfield  {journal} {\bibinfo  {journal}
  {Phys. Rev. E}\ }\textbf {\bibinfo {volume} {110}},\ \bibinfo {pages}
  {014106} (\bibinfo {year} {2024})}\BibitemShut {NoStop}%
\bibitem [{\citenamefont {Maslov}\ \emph {et~al.}(1999)\citenamefont {Maslov},
  \citenamefont {Tang},\ and\ \citenamefont {Zhang}}]{PhysRevLett.83.2449}%
  \BibitemOpen
  \bibfield  {author} {\bibinfo {author} {\bibfnamefont {S.}~\bibnamefont
  {Maslov}}, \bibinfo {author} {\bibfnamefont {C.}~\bibnamefont {Tang}}, \ and\
  \bibinfo {author} {\bibfnamefont {Y.-C.}\ \bibnamefont {Zhang}},\ }{\it
  \bibinfo {title} {$1/\mathit{f}$ Noise in Bak-Tang-Wiesenfeld Models on
  Narrow Stripes},}\ \href {\doibase 10.1103/PhysRevLett.83.2449} {\bibfield
  {journal} {\bibinfo  {journal} {Phys. Rev. Lett.}\ }\textbf {\bibinfo
  {volume} {83}},\ \bibinfo {pages} {2449} (\bibinfo {year}
  {1999})}\BibitemShut {NoStop}%
\bibitem [{\citenamefont {Yadav}\ \emph {et~al.}(2012)\citenamefont {Yadav},
  \citenamefont {Ramaswamy},\ and\ \citenamefont {Dhar}}]{PhysRevE.85.061114}%
  \BibitemOpen
  \bibfield  {author} {\bibinfo {author} {\bibfnamefont {A.~C.}\ \bibnamefont
  {Yadav}}, \bibinfo {author} {\bibfnamefont {R.}~\bibnamefont {Ramaswamy}}, \
  and\ \bibinfo {author} {\bibfnamefont {D.}~\bibnamefont {Dhar}},\ }{\it
  \bibinfo {title} {Power spectrum of mass and activity fluctuations in a
  sandpile},}\ \href {\doibase 10.1103/PhysRevE.85.061114} {\bibfield
  {journal} {\bibinfo  {journal} {Phys. Rev. E}\ }\textbf {\bibinfo {volume}
  {85}},\ \bibinfo {pages} {061114} (\bibinfo {year} {2012})}\BibitemShut
  {NoStop}%
\bibitem [{\citenamefont {Yadav}\ and\ \citenamefont
  {Kumar}(2022)}]{Yadav_2022}%
  \BibitemOpen
  \bibfield  {author} {\bibinfo {author} {\bibfnamefont {A.~C.}\ \bibnamefont
  {Yadav}}\ and\ \bibinfo {author} {\bibfnamefont {N.}~\bibnamefont {Kumar}},\
  }{\it \bibinfo {title} {A cutoff time scaling of 1/f noise in a sandpile},}\
  \href {\doibase 10.1209/0295-5075/ac4f09} {\bibfield  {journal} {\bibinfo
  {journal} {EPL}\ }\textbf {\bibinfo {volume} {137}},\ \bibinfo {pages}
  {12003} (\bibinfo {year} {2022})}\BibitemShut {NoStop}%
\bibitem [{\citenamefont {De~Los~Rios}\ and\ \citenamefont
  {Zhang}(1999)}]{PhysRevLett.82.472}%
  \BibitemOpen
  \bibfield  {author} {\bibinfo {author} {\bibfnamefont {P.}~\bibnamefont
  {De~Los~Rios}}\ and\ \bibinfo {author} {\bibfnamefont {Y.-C.}\ \bibnamefont
  {Zhang}},\ }{\it \bibinfo {title} {Universal $1/\mathit{f}$ Noise from
  Dissipative Self-Organized Criticality Models},}\ \href {\doibase
  10.1103/PhysRevLett.82.472} {\bibfield  {journal} {\bibinfo  {journal} {Phys.
  Rev. Lett.}\ }\textbf {\bibinfo {volume} {82}},\ \bibinfo {pages} {472}
  (\bibinfo {year} {1999})}\BibitemShut {NoStop}%
\bibitem [{\citenamefont {Kumar}\ \emph {et~al.}(2022)\citenamefont {Kumar},
  \citenamefont {Singh},\ and\ \citenamefont {Yadav}}]{Kumar_2022}%
  \BibitemOpen
  \bibfield  {author} {\bibinfo {author} {\bibfnamefont {N.}~\bibnamefont
  {Kumar}}, \bibinfo {author} {\bibfnamefont {S.}~\bibnamefont {Singh}}, \ and\
  \bibinfo {author} {\bibfnamefont {A.~C.}\ \bibnamefont {Yadav}},\ }{\it
  \bibinfo {title} {Energy fluctuations in one dimensional Zhang sandpile
  model},}\ \href {\doibase 10.1088/1742-5468/ac7aa8} {\bibfield  {journal}
  {\bibinfo  {journal} {J. Stat. Mech.}\ }\textbf {\bibinfo {volume} {2022}},\
  \bibinfo {pages} {073203} (\bibinfo {year} {2022})}\BibitemShut {NoStop}%
\bibitem [{\citenamefont {Bak}\ and\ \citenamefont
  {Sneppen}(1993)}]{PhysRevLett.71.4083}%
  \BibitemOpen
  \bibfield  {author} {\bibinfo {author} {\bibfnamefont {P.}~\bibnamefont
  {Bak}}\ and\ \bibinfo {author} {\bibfnamefont {K.}~\bibnamefont {Sneppen}},\
  }{\it \bibinfo {title} {Punctuated equilibrium and criticality in a simple
  model of evolution},}\ \href {\doibase 10.1103/PhysRevLett.71.4083}
  {\bibfield  {journal} {\bibinfo  {journal} {Phys. Rev. Lett.}\ }\textbf
  {\bibinfo {volume} {71}},\ \bibinfo {pages} {4083} (\bibinfo {year}
  {1993})}\BibitemShut {NoStop}%
\bibitem [{\citenamefont {Zaitsev}(1992)}]{ZAITSEV1992411}%
  \BibitemOpen
  \bibfield  {author} {\bibinfo {author} {\bibfnamefont {S.~I.}\ \bibnamefont
  {Zaitsev}},\ }{\it \bibinfo {title} {Robin Hood as self-organized
  criticality},}\ \href {\doibase https://doi.org/10.1016/0378-4371(92)90053-S}
  {\bibfield  {journal} {\bibinfo  {journal} {Physica A}\ }\textbf {\bibinfo
  {volume} {189}},\ \bibinfo {pages} {411} (\bibinfo {year}
  {1992})}\BibitemShut {NoStop}%
\bibitem [{\citenamefont {Maslov}\ \emph {et~al.}(1994)\citenamefont {Maslov},
  \citenamefont {Paczuski},\ and\ \citenamefont {Bak}}]{PhysRevLett.73.2162}%
  \BibitemOpen
  \bibfield  {author} {\bibinfo {author} {\bibfnamefont {S.}~\bibnamefont
  {Maslov}}, \bibinfo {author} {\bibfnamefont {M.}~\bibnamefont {Paczuski}}, \
  and\ \bibinfo {author} {\bibfnamefont {P.}~\bibnamefont {Bak}},\ }{\it
  \bibinfo {title} {Avalanches and $\frac{1}{f}$ Noise in Evolution and Growth
  Models},}\ \href {\doibase 10.1103/PhysRevLett.73.2162} {\bibfield  {journal}
  {\bibinfo  {journal} {Phys. Rev. Lett.}\ }\textbf {\bibinfo {volume} {73}},\
  \bibinfo {pages} {2162} (\bibinfo {year} {1994})}\BibitemShut {NoStop}%
\bibitem [{\citenamefont {Paczuski}\ \emph {et~al.}(1996)\citenamefont
  {Paczuski}, \citenamefont {Maslov},\ and\ \citenamefont
  {Bak}}]{PhysRevE.53.414}%
  \BibitemOpen
  \bibfield  {author} {\bibinfo {author} {\bibfnamefont {M.}~\bibnamefont
  {Paczuski}}, \bibinfo {author} {\bibfnamefont {S.}~\bibnamefont {Maslov}}, \
  and\ \bibinfo {author} {\bibfnamefont {P.}~\bibnamefont {Bak}},\ }{\it
  \bibinfo {title} {Avalanche dynamics in evolution, growth, and depinning
  models},}\ \href {\doibase 10.1103/PhysRevE.53.414} {\bibfield  {journal}
  {\bibinfo  {journal} {Phys. Rev. E}\ }\textbf {\bibinfo {volume} {53}},\
  \bibinfo {pages} {414} (\bibinfo {year} {1996})}\BibitemShut {NoStop}%
\bibitem [{\citenamefont {Daerden}\ and\ \citenamefont
  {Vanderzande}(1996)}]{PhysRevE.53.4723}%
  \BibitemOpen
  \bibfield  {author} {\bibinfo {author} {\bibfnamefont {F.}~\bibnamefont
  {Daerden}}\ and\ \bibinfo {author} {\bibfnamefont {C.}~\bibnamefont
  {Vanderzande}},\ }{\it \bibinfo {title} {$\frac{1}{f}$ noise in the
  Bak-Sneppen model},}\ \href {\doibase 10.1103/PhysRevE.53.4723} {\bibfield
  {journal} {\bibinfo  {journal} {Phys. Rev. E}\ }\textbf {\bibinfo {volume}
  {53}},\ \bibinfo {pages} {4723} (\bibinfo {year} {1996})}\BibitemShut
  {NoStop}%
\bibitem [{\citenamefont {Davidsen}\ and\ \citenamefont
  {L\"uthje}(2001)}]{PhysRevE.63.063101}%
  \BibitemOpen
  \bibfield  {author} {\bibinfo {author} {\bibfnamefont {J.}~\bibnamefont
  {Davidsen}}\ and\ \bibinfo {author} {\bibfnamefont {N.}~\bibnamefont
  {L\"uthje}},\ }{\it \bibinfo {title} {Comment on ``$1/f$ noise in the
  Bak-Sneppen model''},}\ \href {\doibase 10.1103/PhysRevE.63.063101}
  {\bibfield  {journal} {\bibinfo  {journal} {Phys. Rev. E}\ }\textbf {\bibinfo
  {volume} {63}},\ \bibinfo {pages} {063101} (\bibinfo {year}
  {2001})}\BibitemShut {NoStop}%
\bibitem [{\citenamefont {Singh}\ \emph {et~al.}(2023)\citenamefont {Singh},
  \citenamefont {Chhimpa},\ and\ \citenamefont {Yadav}}]{PhysRevE.108.044109}%
  \BibitemOpen
  \bibfield  {author} {\bibinfo {author} {\bibfnamefont {A.}~\bibnamefont
  {Singh}}, \bibinfo {author} {\bibfnamefont {R.}~\bibnamefont {Chhimpa}}, \
  and\ \bibinfo {author} {\bibfnamefont {A.~C.}\ \bibnamefont {Yadav}},\ }{\it
  \bibinfo {title} {Fitness fluctuations in the Bak-Sneppen model},}\ \href
  {\doibase 10.1103/PhysRevE.108.044109} {\bibfield  {journal} {\bibinfo
  {journal} {Phys. Rev. E}\ }\textbf {\bibinfo {volume} {108}},\ \bibinfo
  {pages} {044109} (\bibinfo {year} {2023})}\BibitemShut {NoStop}%
\bibitem [{\citenamefont {Boettcher}\ and\ \citenamefont
  {Paczuski}(2000)}]{PhysRevLett.84.2267}%
  \BibitemOpen
  \bibfield  {author} {\bibinfo {author} {\bibfnamefont {S.}~\bibnamefont
  {Boettcher}}\ and\ \bibinfo {author} {\bibfnamefont {M.}~\bibnamefont
  {Paczuski}},\ }{\it \bibinfo {title} {4 Is the Upper Critical Dimension for
  the Bak-Sneppen Model},}\ \href {\doibase 10.1103/PhysRevLett.84.2267}
  {\bibfield  {journal} {\bibinfo  {journal} {Phys. Rev. Lett.}\ }\textbf
  {\bibinfo {volume} {84}},\ \bibinfo {pages} {2267} (\bibinfo {year}
  {2000})}\BibitemShut {NoStop}%
\bibitem [{\citenamefont {Cassidy}(2020)}]{cassidy2020}%
  \BibitemOpen
  \bibfield  {author} {\bibinfo {author} {\bibfnamefont {M.}~\bibnamefont
  {Cassidy}},\ }\href@noop {} {\emph {\bibinfo {title} {Biological Evolution:
  An Introduction}}}\ (\bibinfo  {publisher} {Cambridge University Press,
  Cambridge, England},\ \bibinfo {year} {2020})\BibitemShut {NoStop}%
\bibitem [{\citenamefont {Sanjak}\ \emph {et~al.}(2018)\citenamefont {Sanjak},
  \citenamefont {Sidorenko}, \citenamefont {Robinson}, \citenamefont
  {Thornton},\ and\ \citenamefont {Visscher}}]{sanjak_2018}%
  \BibitemOpen
  \bibfield  {author} {\bibinfo {author} {\bibfnamefont {J.~S.}\ \bibnamefont
  {Sanjak}}, \bibinfo {author} {\bibfnamefont {J.}~\bibnamefont {Sidorenko}},
  \bibinfo {author} {\bibfnamefont {M.~R.}\ \bibnamefont {Robinson}}, \bibinfo
  {author} {\bibfnamefont {K.~R.}\ \bibnamefont {Thornton}}, \ and\ \bibinfo
  {author} {\bibfnamefont {P.~M.}\ \bibnamefont {Visscher}},\ }{\it \bibinfo
  {title} {Evidence of directional and stabilizing selection in contemporary
  humans},}\ \href {\doibase 10.1073/pnas.1707227114} {\bibfield  {journal}
  {\bibinfo  {journal} {Proc. Natl. Acad. Sci. U.S.A.}\ }\textbf {\bibinfo
  {volume} {115}},\ \bibinfo {pages} {151} (\bibinfo {year}
  {2018})}\BibitemShut {NoStop}%
\bibitem [{\citenamefont {Glazov}\ and\ \citenamefont
  {Sherman}(2011)}]{PhysRevLett.107.156602}%
  \BibitemOpen
  \bibfield  {author} {\bibinfo {author} {\bibfnamefont {M.~M.}\ \bibnamefont
  {Glazov}}\ and\ \bibinfo {author} {\bibfnamefont {E.~Y.}\ \bibnamefont
  {Sherman}},\ }{\it \bibinfo {title} {Theory of Spin Noise in Nanowires},}\
  \href {\doibase 10.1103/PhysRevLett.107.156602} {\bibfield  {journal}
  {\bibinfo  {journal} {Phys. Rev. Lett.}\ }\textbf {\bibinfo {volume} {107}},\
  \bibinfo {pages} {156602} (\bibinfo {year} {2011})}\BibitemShut {NoStop}%
\end{thebibliography}%
\bibliographystyle{myrev}

\end{document}